\documentclass[preprint2,openany]{aastex7}

\shorttitle{GRB evolution aided by ML redshifts}
\shortauthors{Bal et al.}

\usepackage{graphicx}
\usepackage{float}
\usepackage{amsmath}
\usepackage{subfigure}
\usepackage{xcolor}
\usepackage{makecell}
\usepackage{multirow}
\usepackage{soul}
\usepackage{url}
\setstcolor{red}

\begin{document}

\title{Probing evolution of Long GRB properties through their cosmic formation\\ history aided by Machine Learning predicted redshifts}

\author[0009-0003-8022-8151]{Dhruv S. Bal}\email{dbal@clemson.edu}\thanks{Email: dbal@clemson.edu}
\affiliation{Department of Physics and Astronomy, Clemson University, Clemson, SC 29634, USA}

\author[0000-0003-1265-2981]{Aditya Narendra}\email{narendraaditya8@gmail.com}
\affiliation{Doctoral School of Exact and Natural Sciences, Jagiellonian University, Kraków, Poland}
\affiliation{Astronomical Observatory of Jagiellonian University, Kraków, Poland}

\author[0000-0003-4442-8546]{Maria Giovanna Dainotti}\email{maria.dainotti@nao.ac.jp}
\affiliation{National Astronomical Observatory of Japan, Mitaka, Tokyo 181-8588, Japan}
\affiliation{The Graduate University for Advanced Studies, SOKENDAI, Kanagawa 240-0193, Japan}
\affiliation{Space Science Institute, Boulder, CO 80301, USA}
\affiliation{Center for Astrophysics, University of Nevada, Las Vegas, NV 89154, USA}
\affiliation{Bay Environmental Institute, P.O. Box 25, Moffett Field, CA 94035, USA}

\author[0009-0002-2068-3411]{Nikita S. Khatiya}\email{nkhatiy@clemson.edu}
\affiliation{Department of Physics and Astronomy, Clemson University, Clemson, SC 29634, USA}

\author[0000-0003-1943-010X]{Aleksander Ł. Lenart}\email{aleksander.lenart@student.uj.edu.pl}
\affiliation{Doctoral School of Exact and Natural Sciences, Jagiellonian University, Kraków, Poland}
\affiliation{Astronomical Observatory of Jagiellonian University, Kraków, Poland}

\author[0000-0002-8028-0991]{Dieter H. Hartmann}\email{hdieter@clemson.edu}
\affiliation{Department of Physics and Astronomy, Clemson University, Clemson, SC 29634, USA}

\begin{abstract}
Gamma-ray Bursts (GRBs) are valuable probes of cosmic star formation reaching back into the epoch of reionization, and a large dataset with known redshifts ($z$) is an important ingredient for these studies. Usually, $z$  is measured using spectroscopy or photometry, but $\sim$80\% of GRBs lack such data. Prompt and afterglow correlations can provide estimates in these cases, though they suffer from systematic uncertainties due to assumed cosmologies and due to detector threshold limits. We use a sample with $z$ estimated via machine learning models, based on prompt and afterglow parameters, without relying on cosmological assumptions. We then use an augmented sample of GRBs with measured and predicted redshifts, forming a larger dataset.
We find that the predicted redshifts are a crucial step forward in understanding the evolution of GRB properties. We test three cases: no evolution, an evolution of the beaming factor, and an evolution of all terms captured by an evolution factor $(1+z)^\delta$. We find that these cases can explain the density rate in the redshift range between 1-2, but neither of the cases can explain the derived rate densities at smaller and higher redshifts, which may point towards an evolution term different than a simple power law. Another possibility is that this mismatch is due to the non-homogeneity of the sample, e.g., a non-collapsar origin of some long GRB within the sample. 
    
\end{abstract}

\section{Introduction}
Currently detected to a securely determined redshift of $\sim$9.4 \citep{Cucchiara2011}, Gamma-ray Bursts (GRBs) are the most distant known transients. This detectability is owed to their enormous energy output during the burst \citep{2025JHEAp..4700384L}. This allow us to use them as probes for cosmology and high-energy physics laboratories.

We can also turn the problem on its head by assuming that cosmology is well-known, and thus, any evolution of the GRB's properties has the potential to reveal a wealth of information about their progenitors. Several studies in the literature have used both approaches, using GRBs as probes \citep{2020MNRAS.498.5041L,2024ApJ...967L..30D} and using evolution to probe GRB physics \citep{2024JHEAp..44..323F}.

However, in either case, knowing an accurate redshift for a GRB is one of the most important factors. Oftentimes, estimating the redshifts becomes a challenging task as we rely on the afterglow emission to determine the redshift of GRBs \citep{2025A&A...698A..92N}. However, to capture the afterglow, the initial prompt emission must be quickly followed up by instruments operating in other energy ranges, which contain emission lines to determine $z$ \citep{2023FrASS..1024317L}. 

Quick follow-ups depend on many factors and are not always achievable. This causes many GRBs to remain without redshifts. Moreover, the farther the GRB, the more crucial it becomes to capture the afterglow before it fades below telescope detection limits. Thus, only 12\% of GRBs detected by \emph{Swift} and \emph{Fermi} telescopes have a redshift estimate  \citep{2025A&A...698A..92N}.

Due to these reasons, a method independent of using afterglows is necessary for determining redshifts for most GRBs. In the literature, several studies \citep{2001ApJ...554..643R,2003A&A...407L...1A,2005NCimC..28..647A,2004ApJ...609..935Y,2011MNRAS.418.2202D} utilized internal GRB correlations for this purpose. The main hindrance in directly using these correlations is the circularity argument \citep{2025A&A...698A..92N} and the fact that for a small deviation of the luminosity distance, there is a large deviation of redshift $\Delta z=2$, see \citet{2024ApJ...967L..30D,2024ApJS..271...22D}. Since a specific cosmology is assumed when determining correlations and then used to calculate redshift, the method becomes circular.

With the advent of Machine Learning (ML), we now have an option that does not rely on cosmology-dependent correlations, but on relations among the observed properties. We can indeed use ML to determine the parameters that best predict the redshift of GRBs from observed properties. Studies \citep{2025A&A...698A..92N, 2024ApJ...967L..30D,2024ApJS..271...22D} (hereafter N25, D24a, D24b) use prompt and X-ray or optical afterglow parameters of Long GRBs (LGRBs) to train the ML model. 

LGRBs in N25 are classified as GRBs with a duration ($\mathrm{T_{90}}$) greater than 2 seconds as per the canonical definition \citep{1993ApJ...413L.101K}. The motivation for using LGRBs is due to their association with collapsars in the majority of cases \citep{1993ApJ...405..273W,2003Natur.423..847H,2006ARA&A..44..507W,2013ApJ...776...98X} and their connection to the star formation rate density (SFRD) (e.g. \citet{2006ApJ...642..371K,2006ApJ...651..142H, Pescalli:2015yva}). Moreover, N25 uses X-ray plateau correlations as they have less intrinsic scatter than prompt correlations. A detailed description of the X-ray plateau in the GRB afterglow is highlighted in \citet{2020ApJ...904...97D,2025JHEAp..4700384L}. 

In this study, we use the LGRB sample with redshifts predicted by N25 to determine the rate density of LGRBs (LGRB-RD). We then compare this rate with a sample of GRBs with known redshifts and X-ray plateaus. This is called the observed sample (OS) in our paper. The rate density for this sample is determined in \citet{2025ApJ...990...69K} (hereafter K25). We compare our results with the K25 results, SFRD and the LGRB rate density models in the literature. K25 has a detailed discussion on how different LGRB-RD in the literature compare with each other. For a quick glance, please refer to Table 3 of K25 in the Appendix. 

In summary, several studies indicate that LGRB-RD follows the SFRD till $z\sim 6$, but the picture beyond this redshift is unclear \citep{2006ApJ...642..371K,2006ApJ...651..142H, Pescalli:2015yva}. Other studies have shown that LGRBs are biased tracers of SFRD but indicate an excess at low-$z$ or high-$z$ \citep{2002ApJ...574..554L,2004ApJ...609..935Y,2004NewAR..48..237H,2007JCAP...07..003G,2008ApJ...673L.119K,2015ApJS..218...13Y,2015ApJ...806...44P,2024ApJ...963L..12P}. The authors of these studies attribute this to the evolution of GRBs and their progenitor properties, like beaming angle evolution, metallicity evolution, or a different/separate class of GRBs with lower luminosities at low-$z$. 

On the other hand, the recent James Webb Space Telescope (JWST) observations of early galaxies also challenged the traditional view on the high-$z$ SFR. \cite{Harikane2025} detected a high-$z$ excess of SFR measured with JWST. \cite{Matsumoto2024} discusses that GRBs can match those observations and be good tracers of this redshift regime. Indeed, the previous galaxy-observations-based estimates could be biased by the instrument's insensitivity to red, dim, compact galaxies. This highlights the importance of GRBs in future high-$z$ studies on SFR.

The overall idea is that we study the impact that the GRBs with redshifts derived from ML have on the LGRB-RD. In section 2, we describe the data selection, cuts, and the methodology to derive the LGRB-RD. Section 3 includes the results from our study as well as previous studies that follow a similar method. Section 4 explains the methodology for deriving the ``theoretical" rate density. Finally, section 5 describes the conclusions and discussions in relation to the previous results in the literature and the differences/similarities with SFRD. We use the cosmic concordance cosmology throughout the paper: $ \mathrm{H_0}$ = 70 $\mathrm{km\ s^{-1}\ Mpc^{-1}}$, $\Omega_m$ = 0.3, and $\Omega_{\Lambda}$ = 0.7 \citep{2014ARA&A..52..415M}.

\section{Data Selection and Methodology}
\subsection{Data Selection}
We select GRBs detected by the Neil Gehrels Swift Observatory \citep{2005SSRv..120..143B}. Our sample is initially divided into two groups. The first group contains GRBs with X-ray plateaus and known spectroscopic redshifts. There are 242 such GRBs \citep{2025A&A...698A..92N} in the observed sample. These GRBs are used to train the Machine Learning model described in  N25. The authors then use the methodology described in N25 to predict GRB redshifts that have X-ray plateaus but do not have redshift measurements from observations. The paper states that they successfully predict redshifts for 276 GRBs. 

Since these redshifts are predicted by ML, we first check if there are GRBs with high redshift or luminosity errors. To avoid the effects of outliers, we calculate the median of the redshift and luminosity error distribution. Any GRBs that have errors greater than 99$\%$ of the median (roughly $>3\sigma$ away) are removed from the sample. There are 39 GRBs with high redshift errors and 9 GRBs with high luminosity errors. Thus, the predicted sample (PS) after data cuts consists of 228 GRBs. 

The data analysis then proceeds along two paths. The first is the calculation of the LGRB-RD of the predicted sample, and the other is the calculation of the combined sample (CS). The combined sample contains 470 GRBs with a combination of predicted and observed redshifts.

\subsection{Methodology}
We then removed any GRBs that lie below a luminosity threshold due to instrument sensitivity. To determine this threshold, we follow the same steps as those used in D24a. A Kolmogorov-Smirnov (KS) test (see Appendix) between the predicted sample and a sample of all GRBs with X-ray plateaus is used to determine the value of the limiting flux (F$_{lim}$). A similar strategy is employed for the combined sample. 

\begin{figure*}[ht]
    \centering    \includegraphics[width=0.9\columnwidth]{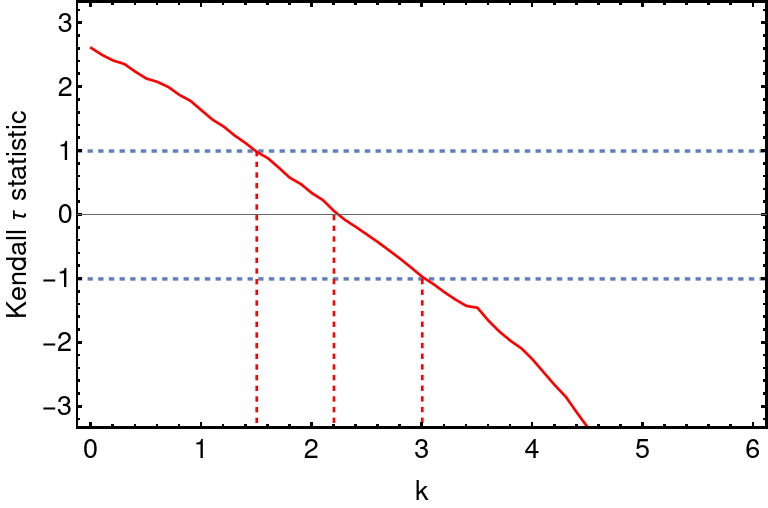}    \includegraphics[width=0.9\columnwidth]{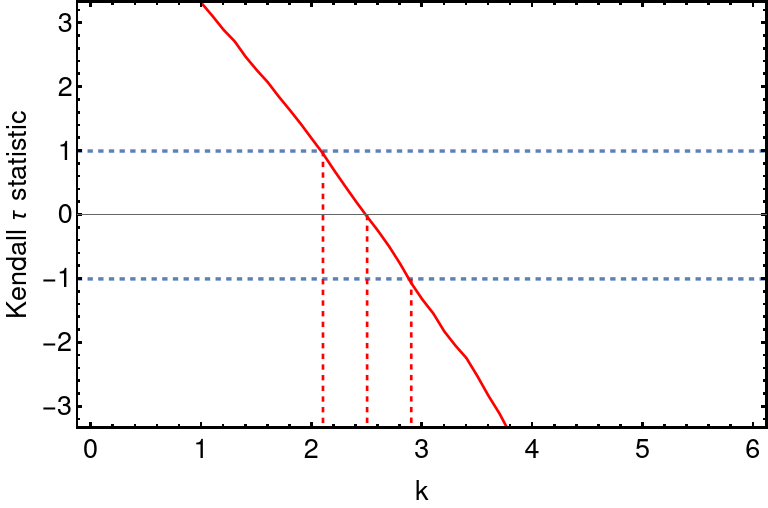}
    \caption{Left: The $\tau$-k distribution for the predicted sample. We select the k value corresponding to $\tau=0$ for making the luminosity and $\mathrm{1+z}$ independent of each other. Right: The $\tau$-k distribution for the combined sample. We select the k value corresponding to $\tau=0$ to make the Luminosity and $1+z$ independent of each other.}
    \label{fig:tau_vs_k}
\end{figure*}

From the KS test, we determine the F$_{lim}$ to be  $2.6 \times 10^{-12} \ \mathrm{erg \  cm^{-2} \ s^{-1}}$ for the predicted sample and $1.4 \times 10^{-12}\ \mathrm{erg \  cm^{-2} \ s^{-1}}$ for the combined sample. A recent study by \citet{2021MNRAS.504.4192B} highlighted the importance of choosing the right F$_{lim}$ and its impact on the results. We note that the overall results remain largely unaffected by the value of F$_{lim}$, as indicated in \citet{2021ApJ...914L..40D}. We discuss this in detail in the Appendix. 

We remove 12 GRBs from the predicted sample as they are below a luminosity threshold, thus reaching 216 GRBs. From the combined sample, we remove 23 GRBs for the same reason, resulting in 447 GRBs.Moreover, we also remove GRB 060614 as this is a long GRB with a merger origin \citep{2015NatCo...6.7323Y,2022Natur.612..228T}. As mentioned in \citet{2025MNRAS.541.3236Z}, there are a couple of additional long GRBs with confirmed merger origins, and a few more that share similar characteristics. We note that none of these GRBs are present in the sample, as they either do not exhibit an X-ray plateau, as seen in the \emph{Swift}-XRT catalog, or do not fit the W07 \citep{2007ApJ...662.1093W} model, or are not present in the \emph{Swift}-XRT catalog and are therefore removed from the sample in N25. The full list of the samples can be found in the GitHub repository\footnote{\url{https://github.com/gammarayapp/GRB-Web-App}}, and the details are described in N25. The final size of the combined sample is 446 GRBs.

Furthermore, due to multiple intrinsic reasons (e.g., metallicity evolution of progenitors), one can expect an ``evolution" of GRB properties such as a typical luminosity. Given that we don't expect a rapid change in properties of possible progenitor systems, we model the $z$-dependent luminosity function as a product of two functions: a luminosity function for GRBs at $z$=0 and the evolutionary function, which is a smooth function of redshift. We thus apply a luminosity correction by assuming a functional form, $\mathrm{L^{\prime}} = \mathrm{L/}(1+z)\mathrm{^k}$, to the two data sets using the Efron-Petrosian (EP) method \citep{1992ApJ...399..345E} following the same steps as explained in D24a, D24b, and K25 and calculate the LGRB-RD. 

\begin{figure*}[ht]
    \centering
    \includegraphics[width=\columnwidth]{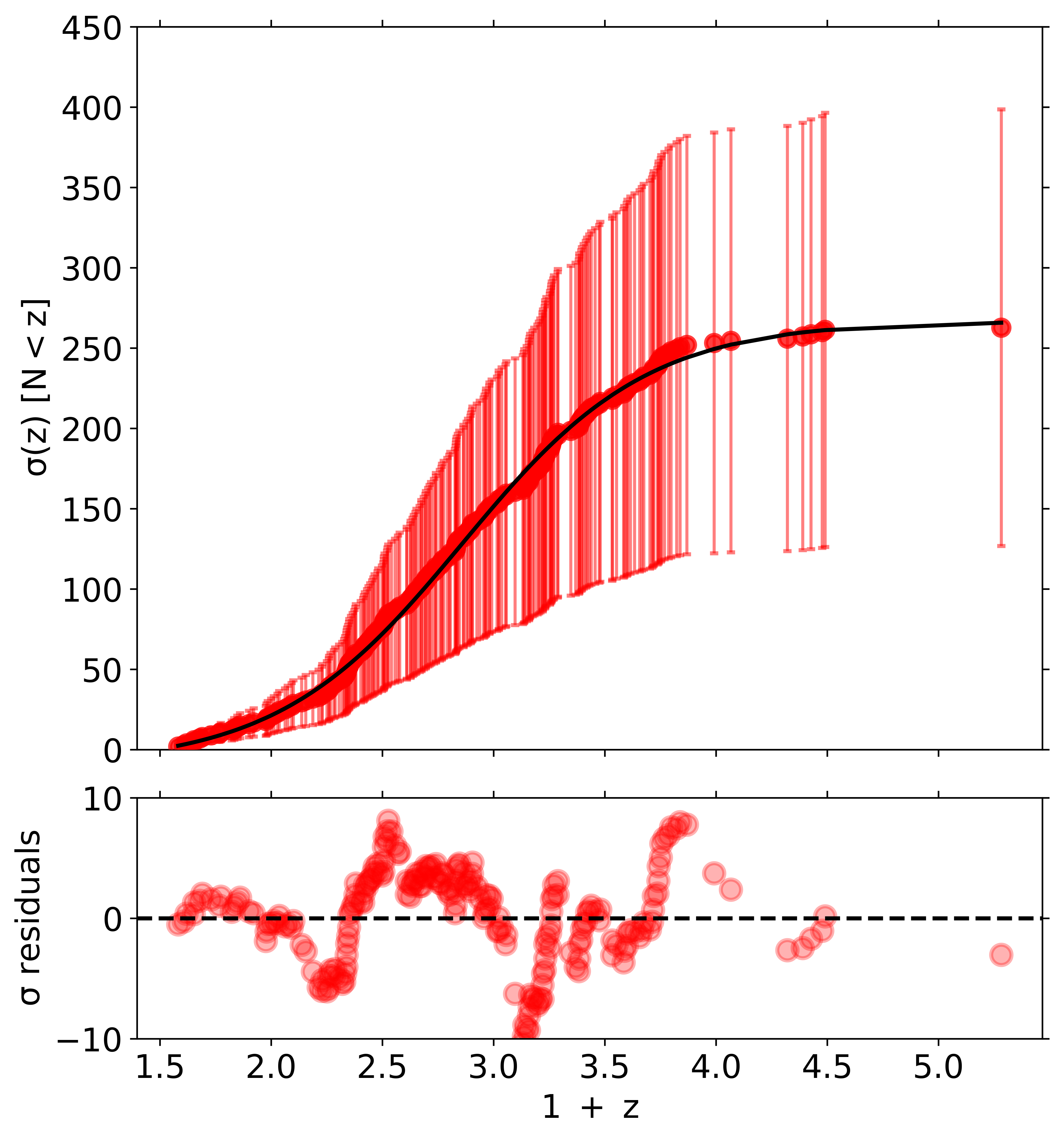}    \includegraphics[width=\columnwidth]{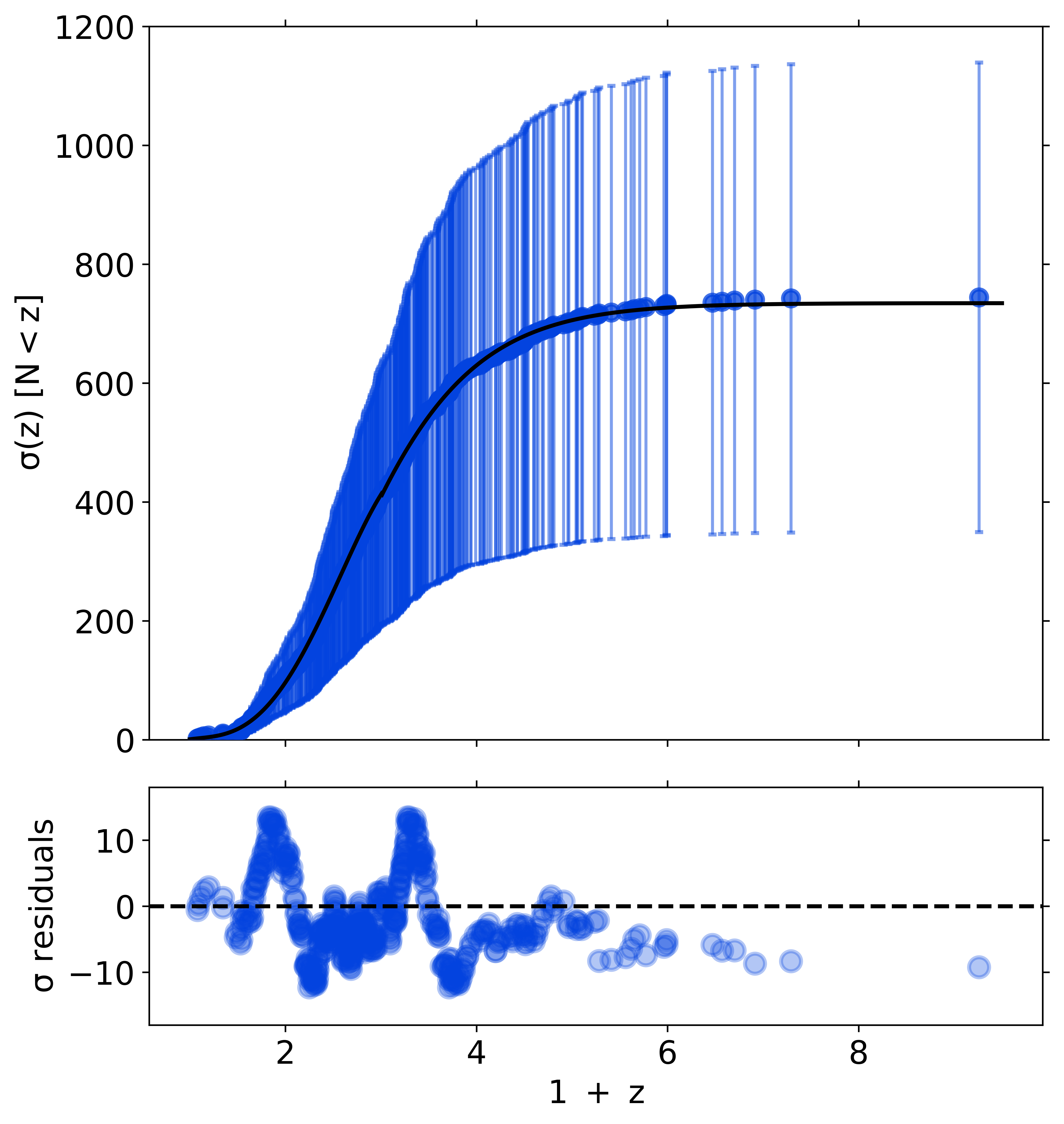}
    \caption{Left panel: The fit to $\sigma$ of the predicted sample with the best parameters derived from the sigmoid function. The best-fit parameters to this function are noted in the first row of Table \ref{tab:sigma_fit_params}. Right panel: The fit to $\sigma$ of the combined sample with the best parameters derived from a piecewise function. The data points below $z_{cut}=3.01$ are best fit with a 4th-order polynomial, while the data points above $z_{cut}$ are best fit with a sigmoid. The best-fit parameters to these functions are noted in the second and third rows of Table \ref{tab:sigma_fit_params}.} 
    \label{fig:sigmafit}
\end{figure*}

\subsubsection{Predicted Sample}\label{sec:PS}
We start by determining the Kendall-$\tau$ statistic for the predicted sample. The statistic is calculated using Equation 3 in K25. The $\tau$-k distribution is shown in the left panel of Figure \ref{fig:tau_vs_k}. The best k value corresponding to a $\tau$ of 0 is $2.2^{+0.8}_{-0.7}$. This corresponds to a significance of $\sim 3\sigma$ from the no evolution case of $k=0$. We then calculate the corrected cumulative rate density distribution ($\sigma_{PS}$, where PS is the predicted sample). $\sigma_{PS}$ is calculated using Equation \ref{eq:sigma}.

\begin{equation}
\label{eq:sigma}
    \sigma(< z_i) = \displaystyle \prod_{j=2}^{i} \left[1 + \frac{1}{M_j}\right].
\end{equation}

The $\sigma_{PS}$ is fit with a function, so that the derivative of the same can be used to calculate the LGRB-RD for the predicted sample. We tested multiple models like Schechter \citep{1976ApJ...203..297S}, several polynomial functions with a maximum degree of 8, a piecewise polynomial, a combination of power-law and Schechter, and a sigmoid \citep{vonSeggern:1993:CSC} to fit $\sigma_{PS}$. The model that best describes the $\sigma$ vs $1+z$ data is the sigmoid. This function is described by Equation \ref{eq:sigmoid}.

\begin{equation}
\label{eq:sigmoid}
    \sigma_{fit} = \frac{a}{1+e^{-s_m(z-z_{b1})}} -c 
\end{equation}

All the models listed above, except the sigmoid, are unable to capture the trend of $\sigma_{PS}$ and miss several data points. Moreover, the $\chi^2$ and similar other metrics are unreliable to determine the best fit function as they assume independent data points and errors. However, we have correlated data points and errors due to the EP method. The best fit is decided based on the residuals and visual fit, and the best fit parameters to $\sigma_{PS}$ are determined with the Levenberg-Marquardt algorithm, and  
given in the first row of Table \ref{tab:sigma_fit_params} while the best fit to $\sigma_{PS}$ and its residuals are shown in the left panel of Figure \ref{fig:sigmafit}.

\begin{deluxetable*}{lccccc}
\tablecaption{$\sigma$ fit parameter results} \label{tab:sigma_fit_params}
\tablehead{
\colhead{Sample and best-fit model } & \colhead{$P_1$} & \colhead{$P_2$} & \colhead{$P_3$} & \colhead{$P_4$} & \colhead{$P_5$} }
\startdata
\multirow{3}{8em}{
PS  \\ 
Sigmoid: $\{a,s_m,z_{b1},c \}$} & & & & & \\
 & 276.5 $\pm$ 4.4 & 2.9 $\pm$ 0.01 & 2.4 $\pm$ 0.04 & 9.9 $\pm$ 0.5 & - \\
 & & & & & \\
 \hline
\multirow{3}{8em}{
CS (z$\leq 3.01$) \\ 
Polynomial: $\{a1,a2,a3,a4,a5 \}$} & & & & & \\
 & -287.3 $\pm$ 54.1 & 829.3 $\pm$ 131.6 & -865.2 $\pm$ 115.9 & 375.5 $\pm$ 43.9 & -51.1 $\pm$ 6.0 \\
 & & & & & \\
 \hline
\multirow{3}{8em}{
CS (z$>3.01$)  \\ 
Sigmoid: $\{a,s_m,z_{b1},c \}$} & & & & & \\
& 1203.3 $\pm$ 258.3 & 2.3 $\pm$ 0.2 & 1.4 $\pm$ 0.07 & 468.7 $\pm$ 256.2 & - \\
 & & & & & \\
\enddata
\tablecomments{Row 1: The first row contains the four best-fit parameters of the sigmoid function that describe the $\sigma_{corr}$ for the predicted sample.  
Row 2: The second row contains the five best-fit parameters of the 4th-order polynomial that describe the $\sigma_{corr}$ for the combined sample below $z_{cut}$ (3.01). Row 3: The third row contains the four best-fit parameters of the sigmoid function that describe the $\sigma_{corr}$ for the combined sample above $z_{cut}$.}
\vspace*{-\baselineskip}
\end{deluxetable*}

\subsubsection{Combined Sample}
We follow the same steps for analyzing the combined sample. The $\tau$-k  distribution for the combined sample can be seen in the right panel of Figure \ref{fig:tau_vs_k}. The best k value corresponding to a $\tau$ of 0 is $2.5^{+0.4}_{-0.4}$. This corresponds to a significance of $\sim 6\sigma$ from the no evolution case of k=0. We further calculate $\sigma_{CS}$ using Equation \ref{eq:sigma} for this sample.

We try the same set of models as described for the predicted sample. We find that the $\sigma_{CS}$ is best fit with a piecewise function, as other models do not accurately capture the trend of the distribution and miss several data points. The part below $z_{cut}$ was fit with a 4th-order polynomial while the part above $z_{cut}$ was fit with the sigmoid function. $z_{cut}$ is considered as 3.01 for this case. The best-fit parameters are listed in the second and third rows of Table \ref{tab:sigma_fit_params} while the fit and its residuals are displayed in the right panel of Figure \ref{fig:sigmafit}. We note that several values of $z_{cut}$ were tested and 3.01 was selected based on the residuals and visual fits. Similar to the predicted sample, we do not use $\chi^2$ and similar other metrics as described in \ref{sec:PS}.

\section{Results}
Similar to the next step in K25 and D24a, we then calculate the LGRB-RD ($\rho$) by taking the derivative of the cumulative distribution function ($\sigma$) for both samples. The $\rho$ is calculated by the following equation

\begin{equation}
    \rho(z) = \frac{d\sigma}{dz} (1+z) \left(\frac{dV}{dz}\right)^{-1}
    \label{rho_z}
\end{equation}

However, this $\rho(z)$ does not account for all the GRBs since we only select GRBs with X-ray plateaus. The true $\rho(z)$ can be found by modifying the rate density to take into account various statistics. In our case, we consider two factors (F$_1$ and F$_2$) accounting for GRBs not included in the analysis by scaling the $\rho(z)$ to the true $\rho(z)$. We describe these factors in detail below. Moreover, we also consider instrument-dependent factors that account for any GRBs missed by the instrument when BAT was not collecting data or was occulted, as seen in Equation \ref{eq: rho_lGRB_z} and motivated by the discussion in \citet{2024ApJ...967L..30D}. 

\begin{equation}
    \rho_{LGRB}(z) = \frac{F_1*F_2* \rho(z),}{Years*Uptime*FOV}
    \label{eq: rho_lGRB_z}
\end{equation}
\begin{figure*}[ht]
    \centering
    \includegraphics[width=\textwidth]{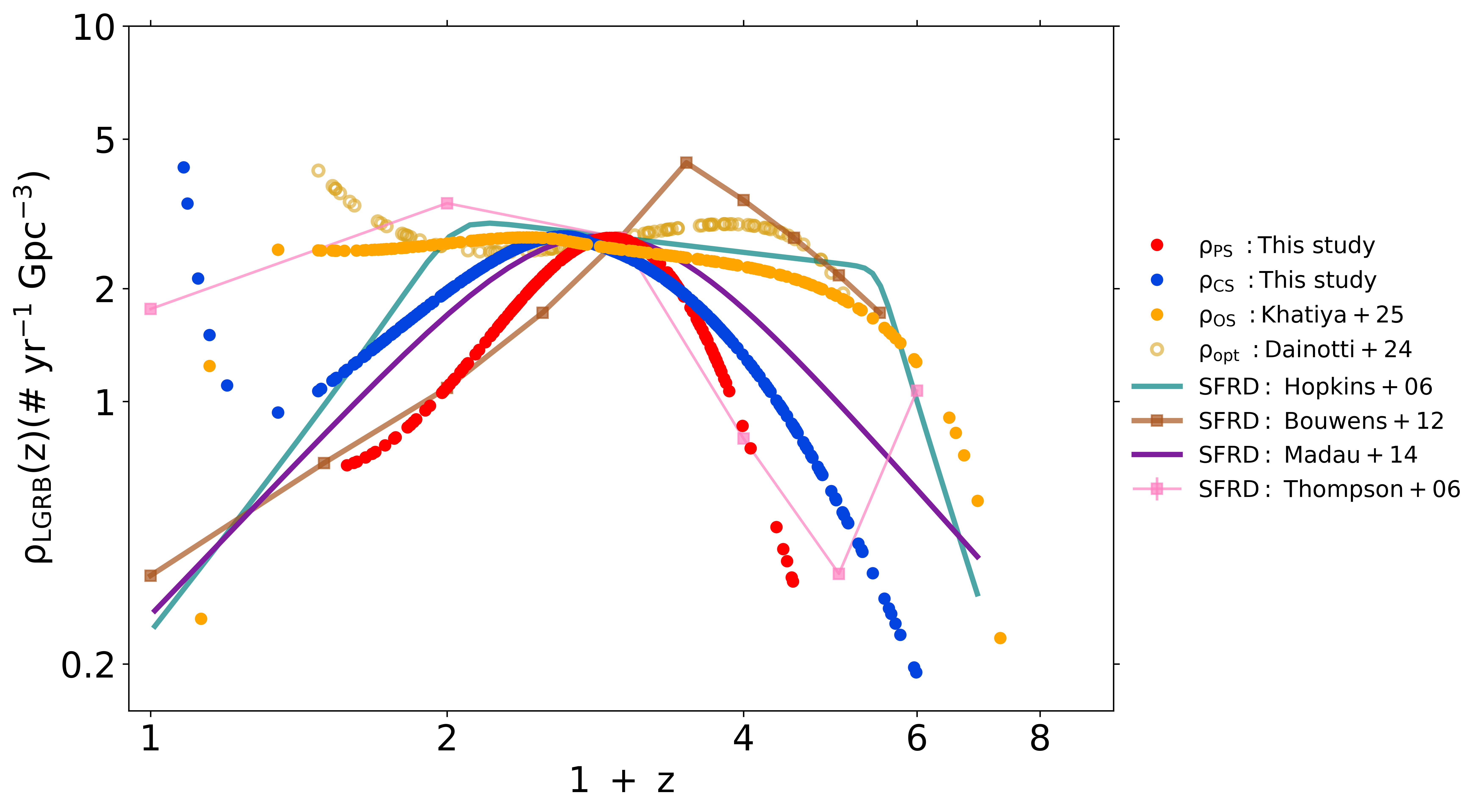}
    \caption{The comparison of $\rho_{PS}(z)$ and $\rho_{CS}(z)$ in this study, with rate densities from the X-ray study of K25 and the optical study of \citet{2024ApJ...967L..30D}. All other data points/lines correspond to various SFRD in the literature as specified in the legend. All curves/data points are renormalized to the predicted sample for better comparison.}
    \label{fig:GRBRD_SFR}
\end{figure*}
Since our sample is constructed from BAT and XRT data, we will consider their operation time, Field of View (FOV), and uptime during the mission. The years of operation are 18.75, as we consider all GRBs till December 10, 2023. The uptime is 75.8\% \citep{2016ApJ...818..110B} and FOV is 0.17 ($\sim 2/12.57$ steradians) \citep{2014ApJ...783...24L}. For both samples, F1 accounts for all GRBs that were excluded from the analysis due to instrument sensitivity. From this definition, F1 = (\# of GRBs before luminosity threshold cut)/(\# of GRBs after luminosity threshold cut). The F2 scaling is applied because we want to consider all 1420 GRBs detected by BAT (until December 10, 2023), regardless of whether they have an X-ray plateau or a redshift estimate. Thus, F2 is defined as F2 =( Total \# of GRBs detected by BAT)/(\# of GRBs before luminosity threshold cut). For the predicted sample, F1$=228/216$ and F2$=1420/228$; while F1$=470/446$ and F2$=1420/470$ for the combined sample.  

After accounting for all the factors, we compare the rate densities obtained from various samples in Figure \ref{fig:GRBRD_SFR}. The figure also includes various SFRD models taken from the literature.

\section{Computing $\rho_{theor}$}

Similar to D24a and K25, we want to compare the rate density derived from SFRD to test whether GRBs are tracers of star formation history. With the assumption that LGRBs occur due to collapsars, any discrepancy between the SFRD and LGRB-RD has the potential to reveal information about GRBs and their progenitor properties. To construct the $\rho_{theor}$ we start with an SFRD ($\mathrm{M_{\odot} \ yr^{-1} \ Mpc^{-3}}$) as described by \citet{2014ARA&A..52..415M} (MD14): 

\begin{equation}
    \mathrm{\rho_{*}(z) = 0.015 \frac{ (1+z)^{2.7}}{1 + (\frac{1 + z}{2.9})^{5.6}}}
\end{equation}

The SFRD is then multiplied by several factors which are taken into account for ``conversion" of a star to a GRB. Equation \ref{eq: Drake_eqn} describes the various factors involved in this conversion. The equation looks similar to the ``Drake equation" and thus we also call it the Drake equation following \citet {2016ApJ...823..154G} and is consistent with K25.

\begin{multline}
    \rho_{theor}(z) = \rho_{*}(z) \times f_{\mathrm{*/ccSN}} \times f_{\mathrm{ccSN/SN1bc}} \times \\ f_{\mathrm{SN1bc/SN1c}} \times f_{\mathrm{beam}}\times f_{\mathrm{z}}.
    \label{eq: Drake_eqn}
    \end{multline}

We list the various factors in equation \ref{eq: Drake_eqn} but point towards K25 and references therein for a detailed explanation. $f_{\mathrm{*/ccSN}}=4.2 \times 10^{-3} M_{\odot}^{-1}$ \citep{1955ApJ...121..161S}; $\mathrm{f_{H-poor/ccSN}* f_{SN1bc/H-poor}} = 27.8\% * 58.6\% = 16.3\%$ \citep{2020ApJ...904...35P}; $f_{\mathrm{SN1bc/SN1c}}=69\%$ \citep{2016ApJ...823..154G}; $f_{\mathrm{beam}}=1\%$\citep{2022ApJ...929..111F}; $f_{\mathrm{z}}$ accounts for the metallicity for GRB environments as explained by \citet{2016ApJ...823..154G}. We have neglected the factor that accounts for GRBs which are not accompanied by SNe Ib/c \citep{2022ApJ...938...41D}.

\begin{figure*}
\begin{interactive}{js}{grbr_plotly.zip}
\includegraphics[width=0.9\textwidth]{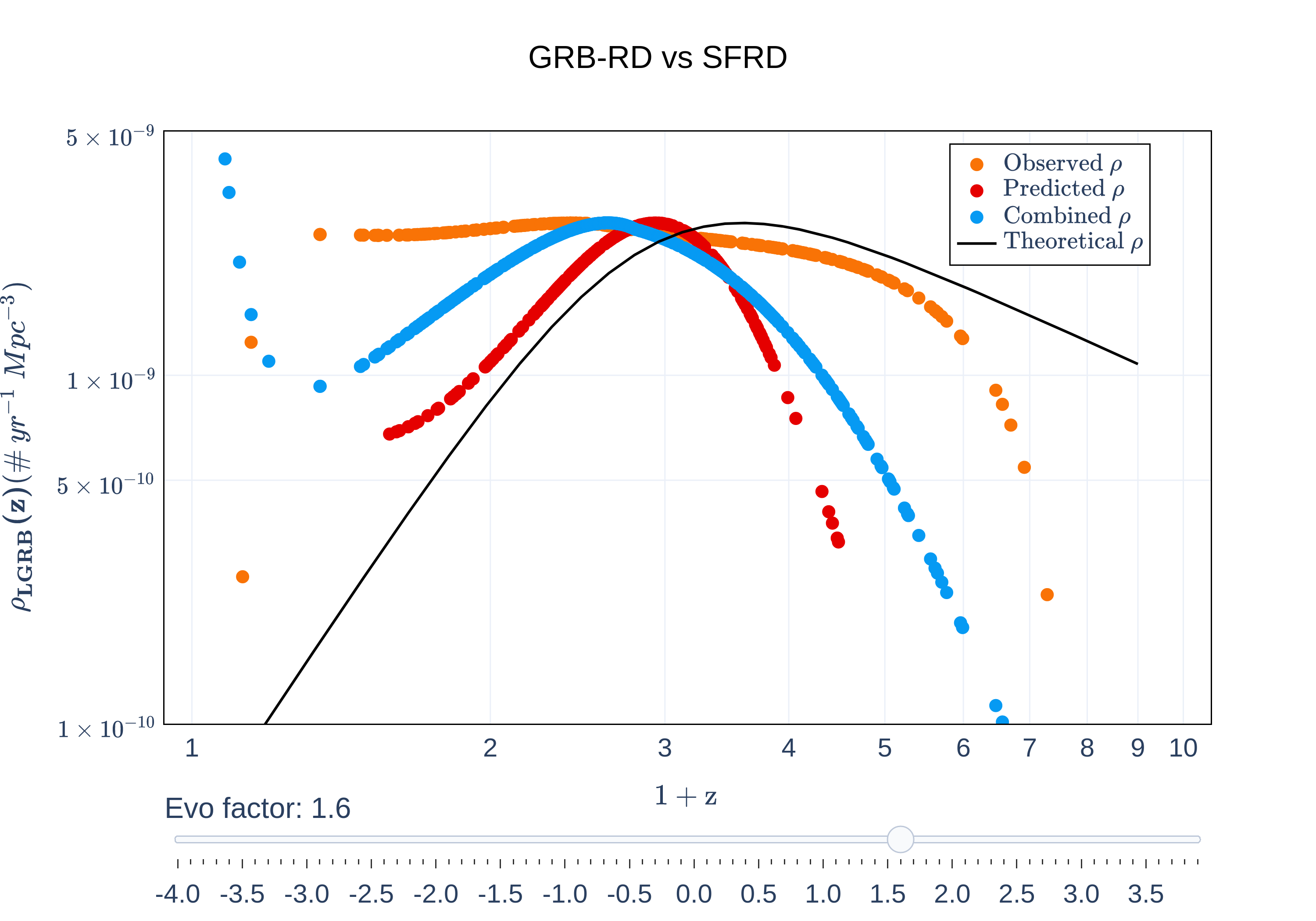}
\end{interactive}
\caption{Example figure of $\rho_{LGRB}(z)$ in 1+z space. The orange-filled circles represent the $\rho_{OS}$ taken from K25. The red-filled and blue-filled circles represent $\rho_{PS}$ and $\rho_{CS}$, respectively. 
The black solid curve is $\rho_{theor}$ derived from Equation \ref{eq: Drake_eqn} and multiplied with $(1+z)^{\delta}$.
We choose $\delta$ = 1.6 as an example for this figure. The interactive figure is available in the online version of the journal, which contains a slider that allows $\delta$ to change, thus accounting for the evolution of GRB/progenitor properties. } 
\label{fig:interactive}
\end{figure*}

Apart from these factors, we also consider another factor that evolves with redshift. For simplicity, we assume this as $(1+z)^\delta$. The equation \ref{eq: Drake_eqn} is also multiplied by this factor. The delta is allowed to vary between -4 and 4. We can also consider this factor to be a beaming evolution factor when testing for a special case of $\delta=1.6$ as found by \citet{2020MNRAS.498.5041L}. For other values of $\delta$, the factor can be considered a mixture of various evolving terms captured into a single evolutionary term.

These cases can be visualized with an interactive figure. Figure \ref{fig:interactive} displays the static image for $\delta=1.6$. Other values can be visualized by moving the slider. The figure contains four elements. The black line is the $\rho_{theor}$. The orange, red, and blue data points are $\rho_{OS}$, $\rho_{PS}$, and $\rho_{CS}$. The points of discussion and the conclusions from the figure are included in the next section.  

\section{Discussion and Conclusion}
Figure \ref{fig:GRBRD_SFR} shows the comparison between $\rho_{LGRB}(z)$ from other samples and the rate density from the optical sample ($\rho_{opt}(z)$). We see that the peak of $\rho_{opt}(z)$ is roughly similar to other rate densities, and they follow a similar trend beyond $z\sim 0.8$. However, an excess is observed at lower redshifts. The difference in shape is mainly due to the fit implemented in the optical sample study. \citet{2024ApJ...967L..30D} fit a polynomial to their cumulative redshift distribution and thus the derivative resulted in the $\rho_{opt}(z)$ seen in Figure \ref{fig:GRBRD_SFR}. This indicates that the derived rate density is sensitive to the fit and derivative at certain redshift regions. Moreover, the rate densities from the three samples show a similar trend. They all have a peak, and the rate density drops on either side of the peak. However, the rate density of the predicted sample peaks at a higher redshift compared to the other two. The peak of the observed sample is not as prominent as the other two samples, and the rate density from the observed sample has a shallow decrease on either side of the peak. The difference in peak is because the training implemented during the machine learning predicts more GRBs towards the mean of the redshift distribution of the observed sample. Thus, the predicted sample was more concentrated near the mean of the observed sample. N25 then applied a bias correction to capture the true trend of the observed sample. Apart from causing the predicted sample to spread out, it also shifted the mean of the sample to a different value. Thus, the two peaks are different in Figure \ref{fig:GRBRD_SFR}. However, the current ML analysis is the most accurate so far, so an additional improvement can be foreseen only when we combine the optical and X-ray samples for ML, thus further enlarging the training set. This is an object of a forthcoming paper.

Finally, an uptick is visible in $\rho_{CS}(z)$ at low-$z$. There are three possible reasons for the presence of this uptick. First, as suggested by \citet{2024ApJ...963L..12P}, the uptick could be due to the contribution of mergers to the rate density. Second, a separate population, such as Low Luminosity GRBs (LLGRBs) \citep{2022MNRAS.513.1078D}, could also cause this uptick. 
Third, the uptick may be due to the fitting technique implemented, and the result could be an artifact of the same. 
An in-depth investigation of this feature will be explored in a future study. The $\rho$'s for the two samples are most reliable in the mid-$z$ range, and we leave the conclusions more open about the low or high-$z$ part of the rate density.

The interactive plot also reveals that the case for no evolution seems to work for the predicted sample. In that, the peaks seem closely aligned, but the $\rho_{PS}(z)$ severely underestimates the rate density derived from SFRD ($\rho_{theor}(z)$). A positive or negative value of $\delta$ can either explain the low-$z$ or high-$z$ end of the predicted sample, but neither value between -4 and 4 captures the entire $\rho_{PS}(z)$. 

An interesting situation emerges for the combined sample. Barring the few low-$z$ points, $\rho_{theor}(z)$ with a $\delta$ of -0.3 follows the $\rho_{CS}(z)$ at low-$z$. While a $\delta$ of -0.9 matches the peak of the rate density and captures most of the high-$z$ end. Moreover, from Figure \ref{fig:GRBRD_SFR}, we see that the rate density of the combined sample peaks at $z\sim1.6$ while the MD14 curve peaks at $z\sim1.9$.  The corresponding lookback time for the peak of the MD14 curve is $\sim 10.1$ Gyr, and the lookback time for the peak of the combined sample is $\sim 9.5^ {+1.4}_{-3.2}$ Gyr. The error on the peak of the lookback time was calculated by averaging the errors of all GRBs between $z$ of 1.6 and 1.9. This difference in both peaks could be attributed to the uncertainty of the predicted redshifts in the combined sample. 

Furthermore, the difference between the lookback times of the peaks is 0.6 Gyr. A massive star of $10\mathrm{M_{\odot}}$ has a lifespan of roughly 0.032 Gyr, see \citet{doi:10.1142/8573} for example. Thus, we have a discrepancy of at least an order of magnitude between the lifetime of a star and the GRB formation rate density, but the lookback time of MD14 lies within the errors of the combined sample. Several authors indicate that the lifespan might increase due to binary interactions \citep{2013ApJ...764..166D,2017A&A...601A..29Z}, which could be another possible reason for the difference between the peaks.

\citet{2022MNRAS.513.1078D} analyzed the full Long GRB sample from \emph{Swift} regardless of whether the GRBs exhibit a plateau in the X-ray afterglow. This is one of the most recent studies with a similar sample and technique for deriving the GRB rate density. They conclude that there is an excess at the low-z end, and the high-z data follow the SFRD closely. These results are similar to the density rate of the combined sample. However, \citet{2022MNRAS.513.1078D} find that the low-z excess is a Gaussian, not a monotonic feature, unlike our results. They discuss the possibility that the low-z population could have different progenitors than the high-z sample, which could also be true in our case. It is likely that the progenitors of GRBs follow SFRD at high-z, whether or not they exhibit a plateau.

Recent works in the literature \citep{2011MNRAS.417.3025V,2019MNRAS.488.4607L} found a good compatibility of GRB rate density and SFR in most of the redshift range, using small, but high-redshift complete samples. However, they present a high-$z$ excess of GRB rate density. This feature is not found in our work. We speculate that this result may be due to multiple factors. The observational biases may play a role; the \emph{Swift}-BAT catalog becomes less redshift-complete with each year. Moreover, it is sensitive to a wide range of luminosities; therefore, we are at a high risk of using a GRB sample with different physical origins. This highlights the importance of the studies on classification and physical interpretation of single progenitor systems. Notably, the GRBs do not occur in typical galaxies. Studies found that LGRBs burst in more compact galaxies - relatively small with high SFR \citep{2022A&A...666A..14S}. Additionally, we are rarely capable of observing hosts further than z=5 \citep{Tanvir2009Nature,Cucchiara2011}. This suggests that such galaxies are unusually dim, an unexpected feature, given that LGRBs are indeed associated with high SFR - high brightness. Recently, JWST found the first two GRB hosts. In both cases, it is a system of two interacting galaxies \citep{2025MNRAS.540.1844T}. Thus, acquiring more such high-$z$ data might lead to better constraints on the nature of GRB host galaxies and GRB rate density.

It is important to note that recent observations have indicated that short GRBs can occur from collapsars and long GRBs can occur from mergers. However, since this study is using the traditional classification of long-short GRBs (boundary at $T_{90}=2$ sec), contamination from the latter case or missing GRBs from the prior case remain a possibility. Additionally, since merger-origin GRBs are affected by a delay-time distribution, this could be another factor contributing to the mismatch between the GRB rate density and the SFRD if the GRB sample is contaminated. However, a follow-up study using a more robust classification scheme to construct the GRB samples may provide further insights into the link between the GRB rate density and the star formation rate density.

In the current scenario, to better constrain the GRB rate density and improve the quality of the $z$-distribution in the predicted sample, a larger training dataset for the ML models will be beneficial. One way to achieve this is by combining the GRBs with measured redshifts having optical plateaus \citep{2024ApJ...967L..30D} and X-ray plateaus \citep{2025A&A...698A..92N}, which will be the focus of a forthcoming paper. However, one must combine the two samples with caution, as the nature of multiwavelength emission in the GRB afterglows is still unknown. For example, \citet{2022MmSAI..93b.132S,2023A&A...675A.117R} found that in some cases the optical and X-ray plateau emission emerges from different emission sites. Moreover, one expects the optical emission to be heavily impacted by the host dust content. Given the redshift evolution of galaxies' dustiness, the GRB formation rate from the optical sample can be significantly biased by the extinction. Therefore, it is crucial to study the rate density from the X-ray and optical samples, as it can provide valuable information about the nature of GRBs, their host galaxies, and their rate densities.

The second idea for increasing the training sample is to use a larger dataset of known redshifts from new telescopes. Recent missions like Space Variable Objects Monitor (\emph{SVOM}) \citep{2021Galax...9..113B} and the \emph{Vera Rubin Observatory} \citep{2006ASPC..351..103A}, and future planned observatories like \emph{Daksha} \citep{2024ExA....57...23B}, Transient High-Energy Sky and Early Universe Surveyor (\emph{THESEUS}) \citep{2021arXiv210208702A}, and High-z Gamma-ray bursts for Unraveling the Dark Ages Mission (\emph{Hi-Z GUNDAM}) \citep{2024SPIE13093E..20Y}, will be instrumental in increasing the GRB sample. However, rapid follow-ups to capture the afterglow will remain a challenging factor in determining GRB redshifts.

\section{Acknowledgements}

We acknowledge Clemson University for the license support received for the Mathematica software. Numerical computations were in part carried out on Small Parallel Computers at the Center for Computational Astrophysics, National Astronomical Observatory of Japan. M.G.D. acknowledges the support of the JSPS Grant-in-Aid for Scientific Research (KAKENHI) (A), Grant Number JP25H00675. We thank the anonymous referee for their valuable feedback. 

\software{This work benefited from the following software: \textsc{Mathematica} \citep{Mathematica},
\textsc{Scipy} \citep{2020SciPy-NMeth}, \textsc{Matplotlib} \citep{Hunter:2007}, and \textsc{Plotly} \citep{plotly}.}

\bibliography{GRB_citations}{}
\bibliographystyle{aasjournal}

\section{Appendix}
The choice of the limiting flux (F$_{lim}$), which accounts for the instrument sensitivity, has been shown to affect the results of the analysis by \citet{2021MNRAS.504.4192B}. In this section, we test various F$_{lim}$ values to understand how significantly the LGRB-RD changes due to this choice. 

\subsection{Predicted Sample}
\begin{figure*}[ht]
    \centering
    \includegraphics[width=\textwidth]{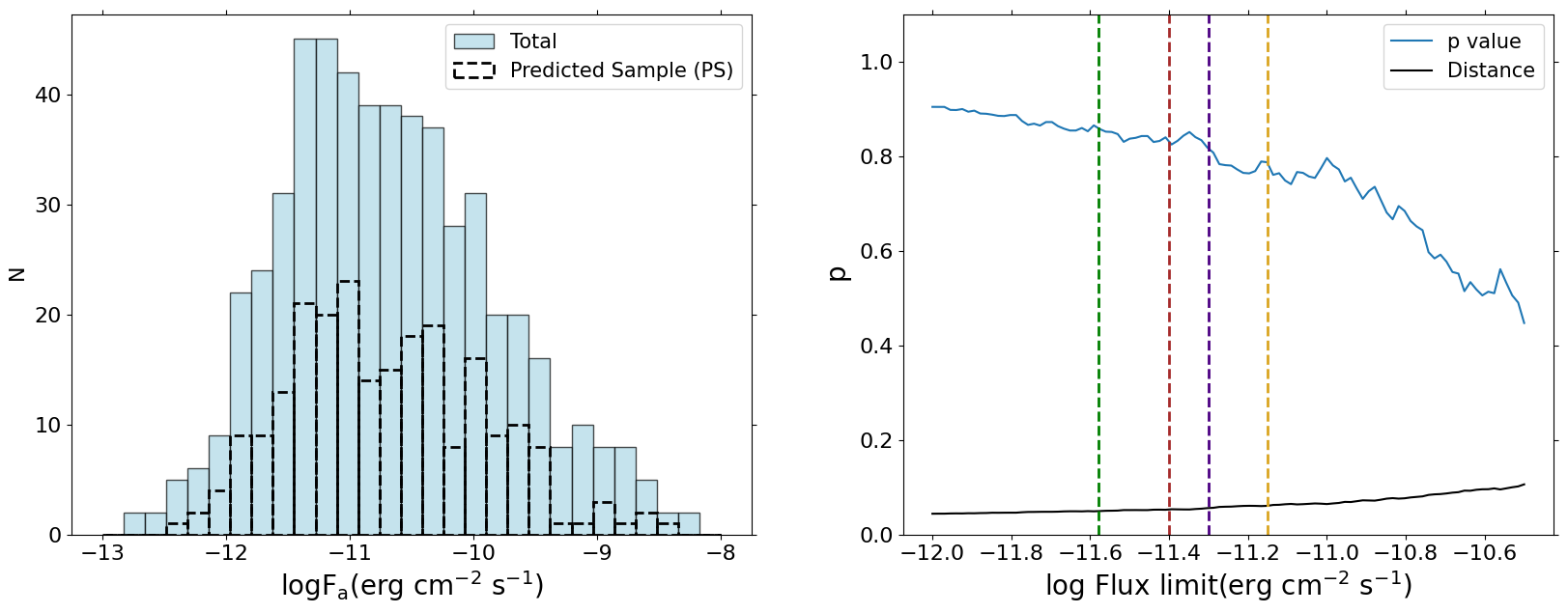}
    \caption{Left panel: The $z$-distribution of the total sample, which contains all GRBs with X-ray plateaus detected by \emph{Swift} and the distribution of the PS. Right panel: The KS test of the two samples in the left panel. We include four cases of $F_{lim}$ corresponding to a 5\%, 10\%, 15\%, and 20\% cut from left to right. The 5\% cut has the highest p-value.}
    \label{fig:KS_test_PS}
\end{figure*}
\begin{figure*}[htbp]
    \centering
    \includegraphics[width=0.45\textwidth]{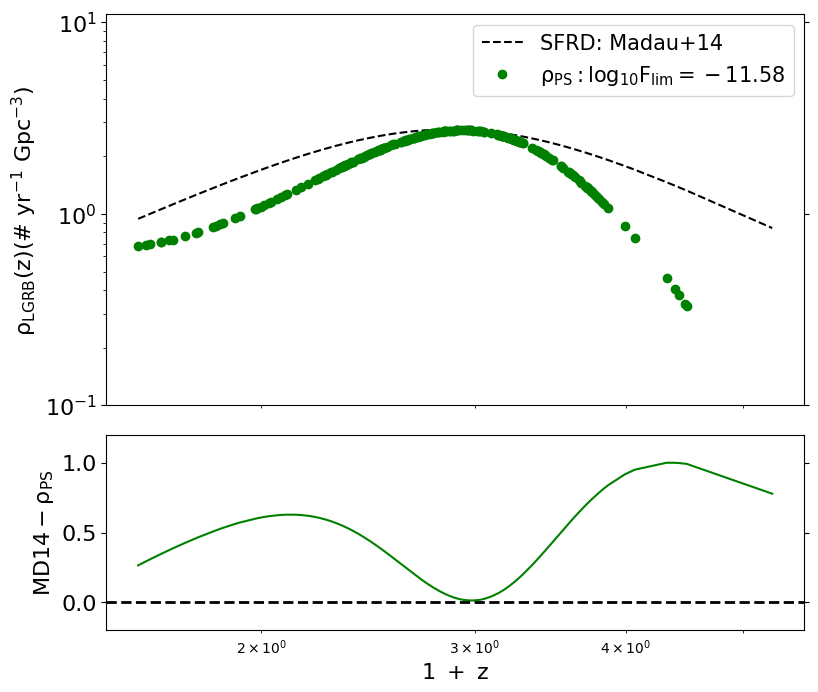}
    \includegraphics[width=0.45\textwidth]{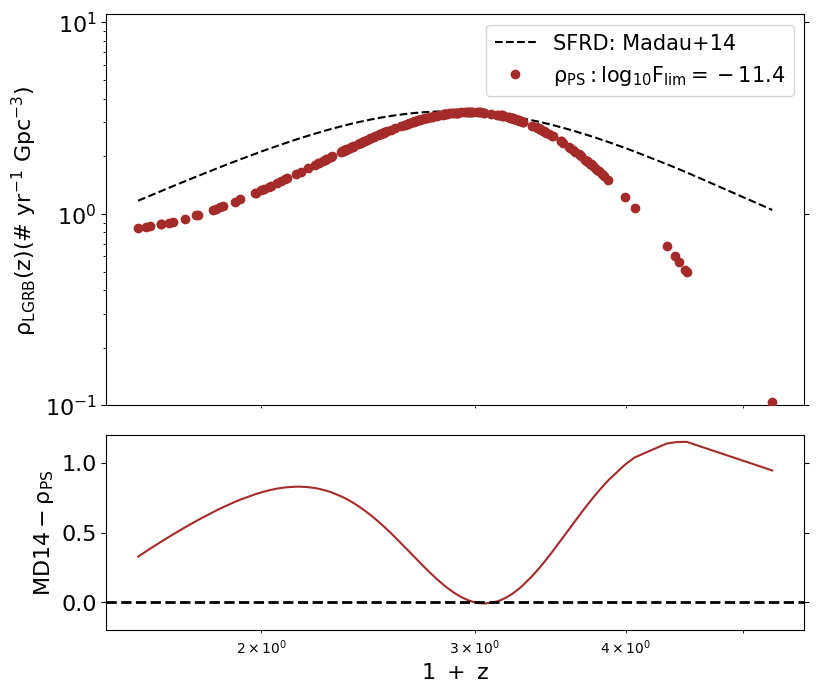}
    \includegraphics[width=0.45\textwidth]{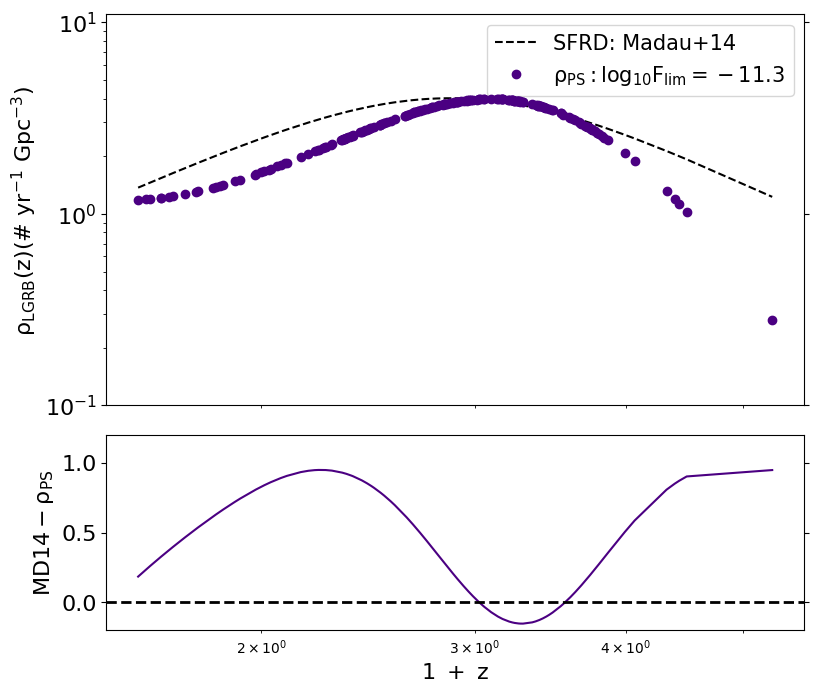}
    \includegraphics[width=0.45\textwidth]{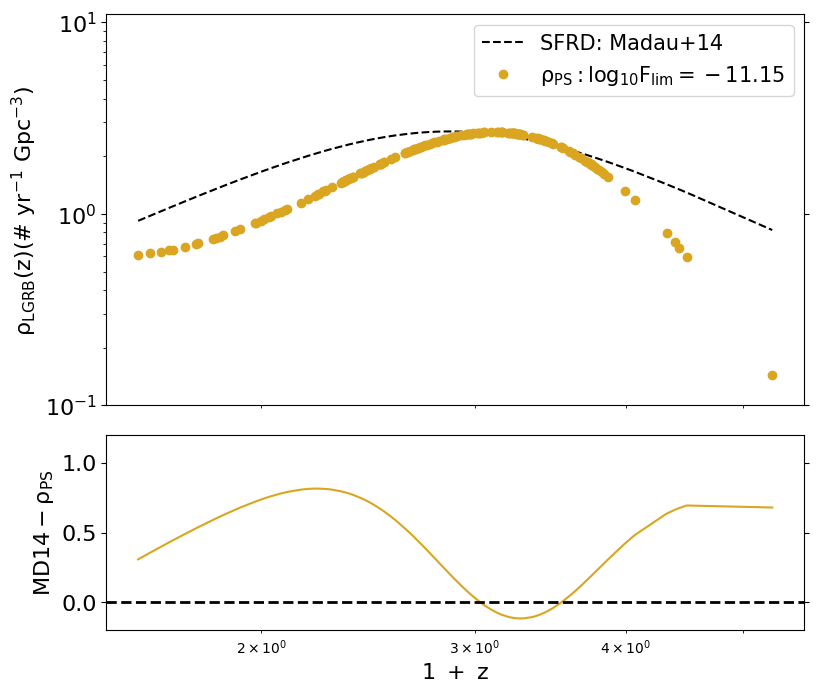}
    \caption{$\rho\mathrm{_{PS}(z)}$ for four flux threshold cases of -11.58, -11.4, -11.3, and -11.15. The filled circles indicate the LGRB-RD corrected for intrinsic luminosity evolution for the PS. The filled circles are color-coded with respect to the vertical dashed lines in the right panel of Figure \ref{fig:KS_test_PS}. The black dashed lines show the SFRD from \citet{2014ARA&A..52..415M} study. The peak of SFRD is renormalized to the peak of LGRB-RD for easier comparison.}
    \label{fig:rho_Flim_PS}
\end{figure*}

Following the methodology outlined in \citet{2021ApJ...914L..40D} and K25, we first perform a KS test between the PS and the sample of all GRBs with an X-ray plateau. We exclude GRBs below a certain flux for both samples and test how similar the two are. Thus, the KS test informs us about the completeness of a sample for a chosen $F_{lim}$. The right panel of Figure \ref{fig:KS_test_PS} shows the result of the KS test for the PS. Any p-value higher than 0.68 indicates that the sample is complete. The figure includes four cases, which reduces the sample by 5\%, 10\%, 15\%, and 20\%. Figure \ref{fig:rho_Flim_PS} shows the rate density for the four cases, and we subtract the MD14 from the rate density to show the magnitude of change. As can be seen from the bottom panels of Figure \ref{fig:rho_Flim_PS} the change is minimal and thus the choice of $F_{lim}$ does not significantly affect the rate density. Thus, we select the 5\% case as it has the highest p-value and the corresponding $F_{lim}$ ($2.6 \times 10^{-12} \ \mathrm{erg \  cm^{-2} \ s^{-1}}$).

\subsection{Combined Sample}

\begin{figure*}[ht]
    \centering
    \includegraphics[width=\textwidth]{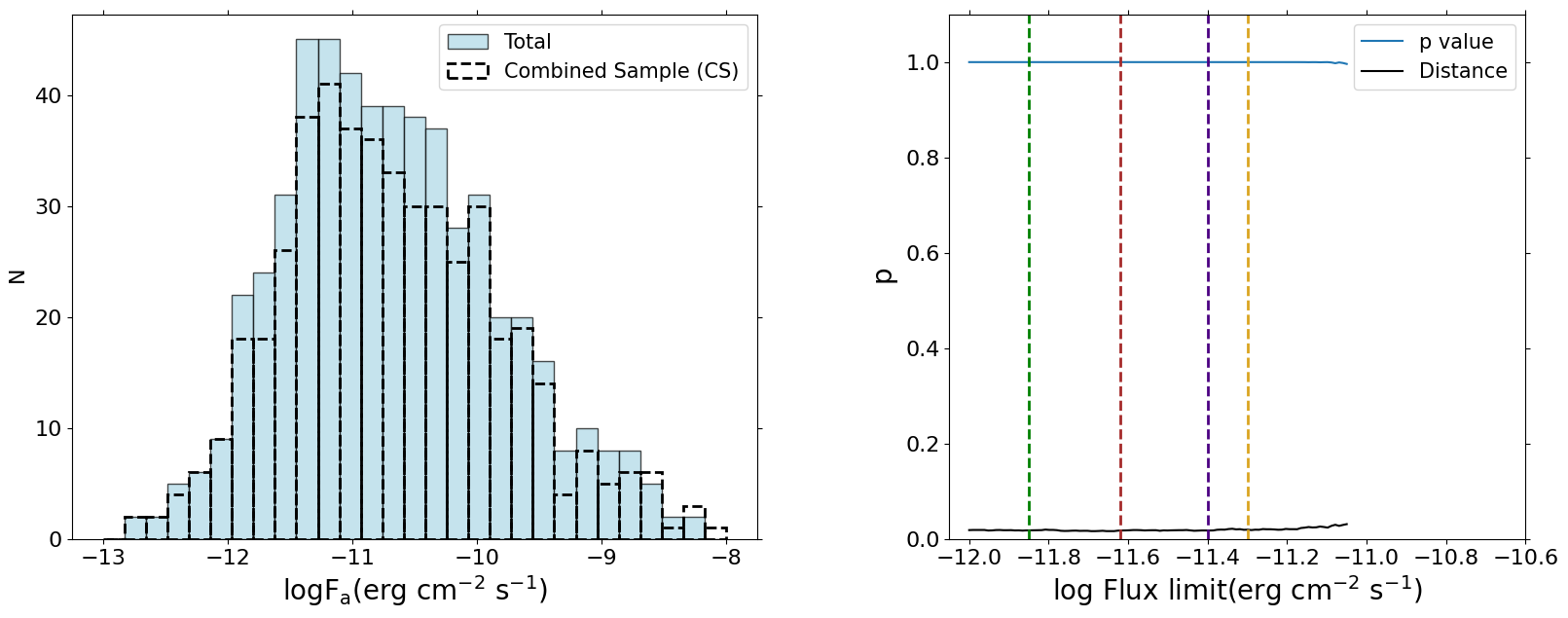}
    \caption{Left panel: The $z$-distribution of the total sample, which contains all GRBs with X-ray plateaus detected by \emph{Swift} and the distribution of the CS. Right panel: The KS test of the two samples in the left panel. We include four cases of $F_{lim}$ corresponding to a 5\%, 10\%, 15\%, and 20\% cut from left to right.}
    \label{fig:KS_test_CS}
\end{figure*}
\begin{figure*}[htbp]
    \centering
    \includegraphics[width=0.45\textwidth]{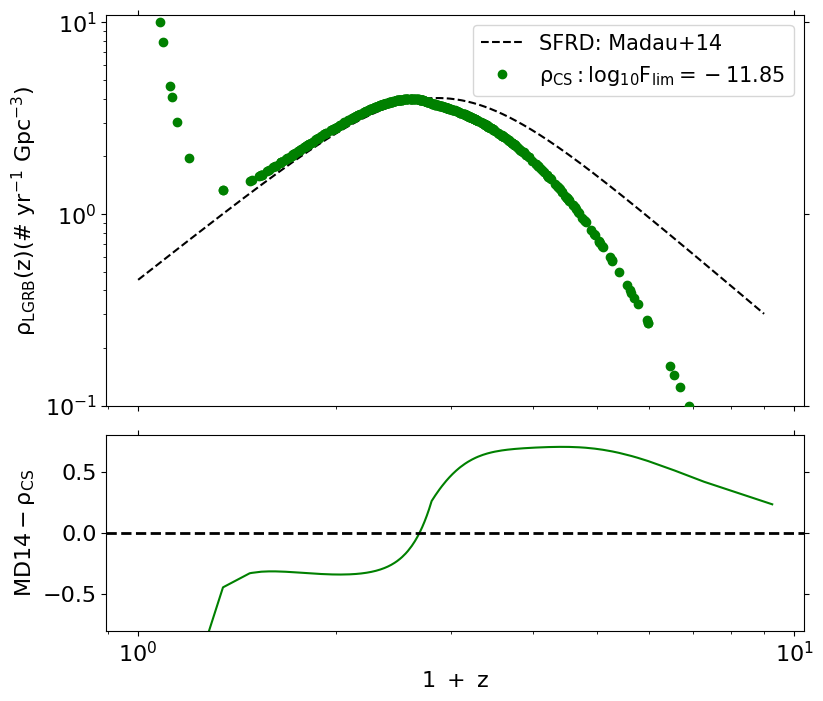}
    \includegraphics[width=0.45\textwidth]{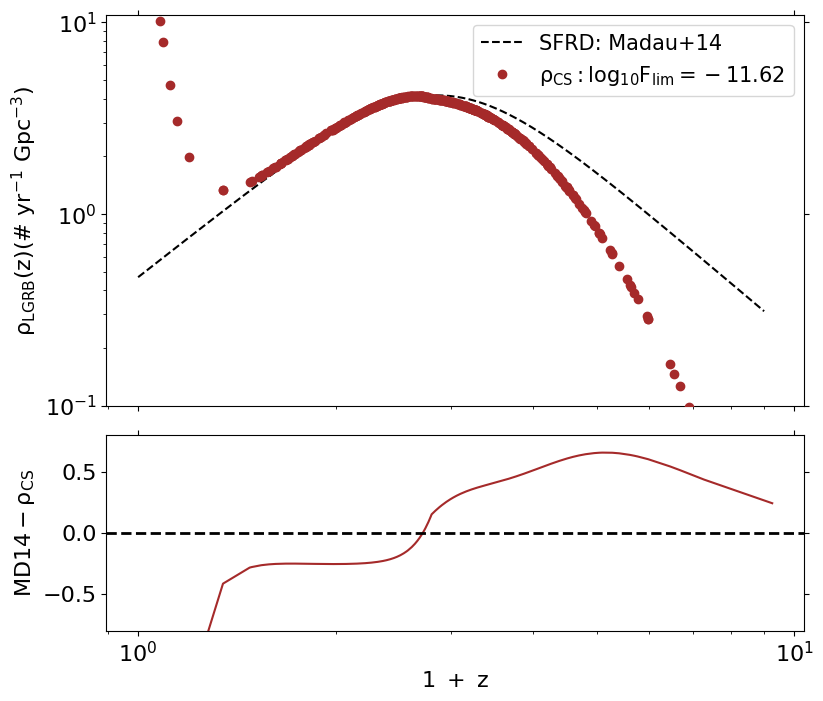}
    \includegraphics[width=0.45\textwidth]{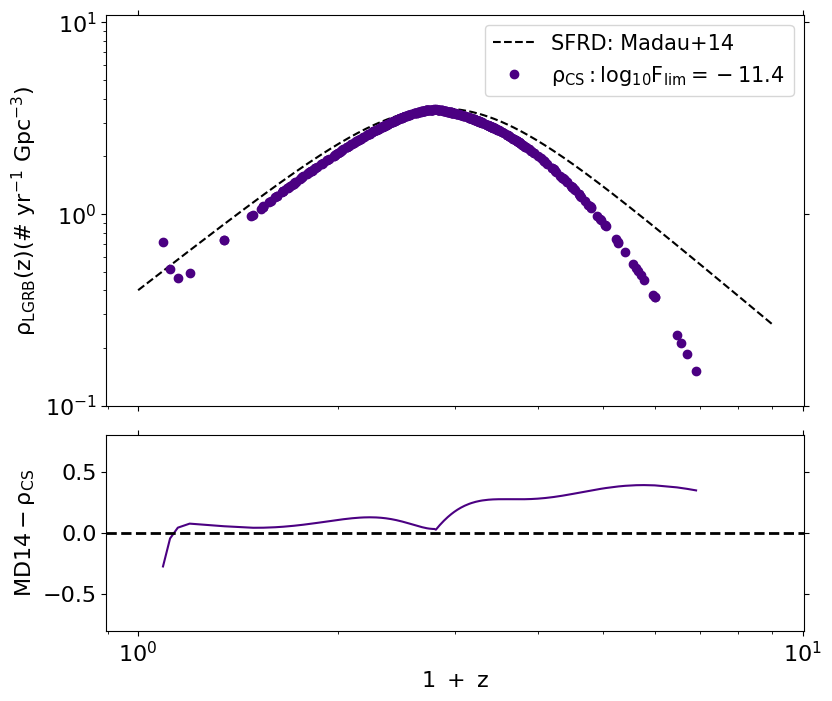}
    \includegraphics[width=0.45\textwidth]{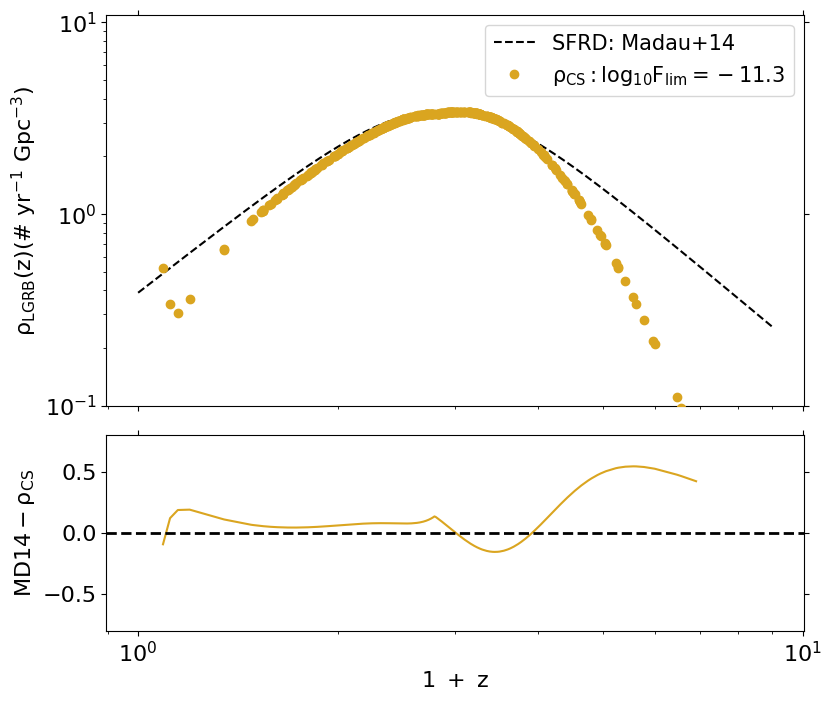}
    \caption{$\rho\mathrm{_{CS}(z)}$ for four flux threshold cases of -11.85, -11.62, -11.4, and -11.3. The filled circles indicate the LGRB-RD corrected for intrinsic luminosity evolution for the CS. The filled circles are color-coded with respect to the vertical dashed lines in the right panel of Figure \ref{fig:KS_test_CS}. The black dashed lines show the SFRD from \citet{2014ARA&A..52..415M} study. The peak of SFRD is renormalized to the peak of LGRB-RD for easier comparison.}
    \label{fig:rho_Flim_CS}
\end{figure*}

We follow a similar method for the CS. The results of the KS are highlighted in the right panel of Figure \ref{fig:KS_test_CS}. As can be seen from the figure, the p-value remains constant ($\sim1$) for most cases of $F_{lim}$, making it difficult to choose an $F_{lim}$ value from the KS test alone. We chose the 5\% case for consistency with the PS and to exclude a reasonable number of GRBs (23) due to instrument sensitivity. Figure \ref{fig:rho_Flim_CS} consists of the rate densities for the four cases. It is clear that the choice of $F_{lim}$ does not affect the mid-$z$ and high-$z$ range significantly. However, the low-$z$ rate and its trend do change based on the choice. In the main text, we offer three possible explanations for low-$z$ trend and are planning an in-depth investigation in a future study.

\end{document}